\documentclass[fleqn]{wlscirep}

\usepackage{graphicx}
\usepackage{xcolor}

\usepackage[utf8]{inputenc}
\usepackage[T1]{fontenc}

\title{Topological Transitions, Pinning and Ratchets for Driven Magnetic Hopfions in Nanostructures}

\author[1, *]{J. C. Bellizotti Souza}
\author[2]{C. J. O. Reichhardt}
\author[2]{C. Reichhardt}
\author[2]{A. Saxena}
\author[3]{N. P. Vizarim}
\author[4]{P. A. Venegas}

\affil[*]{\href{mailto:jc.souza@unesp.br}{jc.souza@unesp.br}}

\affil[1]{POSMAT - Programa de P\'os-Gradua\c{c}\~ao em Ci\^encia e Tecnologia de Materiais, S\~ao Paulo State University (UNESP), School of Sciences, Bauru 17033-360, SP, Brazil}

\affil[2]{Theoretical Division and Center for Nonlinear Studies, Los Alamos National Laboratory, Los Alamos, New Mexico 87545, USA}

\affil[3]{“Gleb Wataghin” Institute of Physics, University of Campinas, 13083-859 Campinas, S\~ao Paulo, Brazil}

\affil[4]{Department of Physics, S\~ao Paulo State University (UNESP), School of Sciences, Bauru 17033-360, SP, Brazil}

\date{\today}


\begin{abstract}
Using atomistic simulations, we examine the dynamics of three-dimensional magnetic hopfions interacting with an array of line defects or posts as a function of defect spacing, defect strength, and current. We find a pinned phase, a sliding phase where a hopfion can move through the posts or hurdles by distorting, and a regime where the hopfion becomes compressed and transforms into a toron that is half the size of the hopfion and moves at a lower velocity. The toron states occur when the defects are strong; however, in the toron regime, it is possible to stabilize sliding hopfions by increasing the applied current. Hopfions move without a Hall angle, while the toron moves with a finite Hall angle. We also show that when a hopfion interacts with an asymmetric array of planar defects, a ratchet effect consisting of a net dc motion can be realized under purely ac driving.
\end{abstract}

\begin{document}
\maketitle

\section{Introduction}
There has been growing interest in particle-like magnetic
textures such as skyrmions, since an increasing number of systems are being found that can support these textured states
\cite{Muhlbauer09,Yu12,Nagaosa13,Woo16,EverschorSitte18}.
In addition to magnetic skyrmions,
a variety of other textures can appear, such as
antiskyrmions \cite{Nayak17a}
skyrmioniums \cite{Kol18}, antiferromagnetic skyrmions \cite{Jani21}, merons \cite{Yu18a},
and biskyrmions \cite{Yu14,Gobel21a}.
These textures can be manipulated or set into motion
with applied currents \cite{Jonietz10,iwasaki_current-induced_2013,Legrand17}
or thermal gradients
\cite{Mochizuki14,Wang20a}.
Moving skyrmions can also interact with pinning sites, and their motion can be controlled by nanostructured patterns
\cite{Juge21,Reichhardt22a}.
Due to their size and stability, magnetic textures
are promising candidates for various applications such as 
memory \cite{Fert17,Finocchio16,Vakili21} and novel computing approaches
\cite{Pinna20,Song20}.
These systems are also interesting in terms of basic science as they represent another particle-like system that can exhibit
depinning phenomena \cite{Lin13,Reichhardt22a}, various sliding phases
\cite{Jonietz10,iwasaki_current-induced_2013,Reichhardt15},
melting \cite{Huang20}, and rectification effects \cite{Gobel21,Souza21}.
In some cases, particularly for skyrmions,
the textures have a strong non-dissipative Magnus
term in their dynamics and move with a Hall angle $\theta$
with respect to an external
drive \cite{Jiang17,Litzius17}.
The Magnus force also affects how such textures
interact with defects in the sample \cite{Zeissler20,Reichhardt22a}.

An example of particle-like textures
that occur in three dimensions is hopfions
\cite{Faddeev76,Faddeev97}, 
which are characterized by their Hopf index \cite{Hopf31,Whitehead47}.
In condensed matter systems, hopfions have been studied
for liquid crystals \cite{Ackerman17,Ackerman17a,Tai18}, 
ferroelectrics \cite{Lukyanchuk20}, and magnetic systems
\cite{Sutcliffe18,Tai18b,Liu18,Kent21,Liu22,Rybakov22,Yu23,Guslienko24,Knapman24,Azhar24}.
Ring type hopfions have recently been experimentally observed
in magnetic systems
\cite{Zheng23}.
Due to their particle-like nature 
and their ability to
be manipulated
with external fields \cite{Wang19,Liu20,Raftrey21,Yu23},
hopfions are another promising system for applications,
and could exhibit new kinds of
dynamical effects when interacting with quenched disorder.
Almost nothing is known, however, about how hopfions interact
with quenched disorder or nanostructured arrays, how stable
they are in such a system, and whether they can
exhibit pinning and depinning behaviors
or undergo topological
transitions when driven through a nanostructured defect array.

In this work, we consider atomistic simulations of three-dimensional (3D)
magnetic hopfions interacting with a comb-like array of line defects
where the hopfions are driven through the defects under an applied current.
We vary the spacing between the defects and modify the defect strengths by changing the local anisotropy. A hopfion subjected to an applied current
moves without a Hall angle. When the hopfion interacts with
the defects, we find a pinned regime where the hopfion cannot move
between the defects, and a sliding regime where
the hopfion can distort and slip between the defects.
For strong disorder or dense defect arrays,
we also find a regime in which the hopfion becomes so strongly compressed as
it passes between the defects that it transforms to a toron that has
a zero Hopf index \cite{Muller20,Liu22,Shimizu24,Amaral25}. 
The toron is half as wide as the hopfion,
moves at a slower velocity, and has a finite Hall angle.
We find that in the regime where a toron forms just above depinning,
if we apply a drive that is well above the depinning threshold,
it is possible to stabilize a moving hopfion since the
distortions experienced by the hopfion from the defect sites
are reduced at higher drives.
We also show that when a hopfion interacts with an asymmetric
array of planar defects under a circular ac drive,
a hopfion ratchet can be realized in which dc motion occurs
where the hopfion translates by one substrate lattice site per ac drive
cycle along the easy direction of the array asymmetry.
Our results indicate that hopfion pinning and motion can
be realized and controlled with nanostructured arrays,
opening the possibility of a variety of applications
that can be both similar to and different from those proposed for skyrmions.

\section{Methods}
Using atomistic simulations, we model a 3D chiral magnet sample of
dimensions 128 nm $\times$ 128 nm$\times$ 17 nm with periodic
boundary conditions along the $x$ and $y$ directions
at zero temperature, $T=0$ K.
The 3D sample is modeled as a cubic arrangement of atoms, with lattice constant
$a=0.5$ nm.
The Hamiltonian governing the atomistic simulations is given by
\cite{evans_atomistic_2018, iwasaki_universal_2013, iwasaki_current-induced_2013}:
\begin{equation}\label{eq:1}
 \mathcal{H}=-\sum_{i, \langle i, j\rangle}J_{ij}{\bf m}_i\cdot{\bf m}_j
 -\sum_{i, \langle i, j\rangle}{\bf D}_{ij}\cdot\left({\bf m}_i\times{\bf m}_j\right)
 -\sum_{i\in V} K_V\left({\bf m}_i\cdot\hat{{\bf z}}\right)^2
 -\sum_{i\in T,B} K_S\left({\bf m}_i\cdot\hat{{\bf z}}\right)^2 .
\end{equation}
The first term is the exchange interaction between first neighbors, with
exchange constant $J_{ij}=J$ between any two magnetic moments. The second
term is the Dzyaloshinskii–Moriya (DM) interaction, where
${\bf D}_{ij}=D{\hat{\bf r}}_{ij}$ is the DM vector, $D$ is
the DM strength, and ${\hat{\bf r}}_{ij}$ is the unit distance vector
between atomic moments $i$ and $j$. This isotropic DM vector
stabilizes Bloch hopfions, which have finite chirality.
The third and fourth terms are perpendicular magnetic anisotropies (PMA),
where $V$ is the sample volume and $T,B$ the sample top and bottom, respectively.
Magnetic hopfions that form in a
chiral magnetic background require confinement in order to remain
stable \cite{Wang19a},
and therefore we impose $K_S\gg K_V$. Although our sample is 17 nm thick,
the confinement reduces the effective thickness to 16 nm.

The time evolution of the magnetic moments is given by \cite{seki_skyrmions_2016, gilbert_phenomenological_2004}:
\begin{equation}\label{eq:2}
 \frac{\text{d} {\bf m}_i}{\text{d} t}=-\gamma{\bf m}_i\times{\bf H}^\text{eff}_i
 +\alpha{\bf m}_i\times\frac{\text{d} {\bf m}_i}{\text{d} t}
 +\frac{j\hbar\gamma P a^3}{2ed\mu}{\bf m}_i\times\left(\hat{{\bf z}}\times\hat{{\bf j}}\right)\times{\bf m}_i \ .
\end{equation}
Here $\gamma$ is the electron gyromagnetic ratio,
${\bf H}^\text{eff}_i=-\frac{1}{\mu}\frac{\partial \mathcal{H}}{\partial {\bf m}_i}$
is the effective field, $\mu=\hbar \gamma$ is the atomic magnetic moment,
$\alpha$ is the phenomenological Gilbert damping,
and the last term is the spin orbit torque (SOT) contribution,
where $j$ is the current density, $P=1$ is the current polarization,
$e$ is the electron charge, $d=17$ nm is the film thickness and
$\hat{{\bf j}}$ is the current direction.

The hopfion is characterized by the Hopf index
\cite{Hopf31,Whitehead47,Faddeev97}, given by:
\begin{equation}\label{eq:3}
 Q_H=\int_V{\bf B}\cdot{\bf A}~\text{d}^3x \\, 
\end{equation}
where the integration is performed over the whole sample.
Here ${\bf B}$ is the emergent magnetic field and
${\bf A}$ is the corresponding vector potential,
${\bf B}=\nabla\times{\bf A}$. The unitless emergent magnetic
field can be computed using 
$ B_i=\frac{1}{8\pi}\varepsilon_{ijk}{\bf m}\cdot\left(\partial_j{\bf m}\times\partial_k{\bf m}\right)$; however, better numerical results are obtained
when using a lattice based approach \cite{Knapman24}, which is the method we
employ throughout this work.
The vector potential ${\bf A}$ used in the Hopf index computation is given by:
\begin{align}
 &A_x(x, y, z)=\int_0^yB_z(x, l, z)~\text{d}l \, , \\
 &A_y(x, y, z)=0 \, , \\
 &A_z(x, y, z)=-\int_0^yB_x(x, l, z)~\text{d}l \, ,
\end{align}
which is known to give good Hopf index computation results \cite{Knapman24}.

We integrate Eq.~\ref{eq:2} using a fourth order Runge-Kutta integration
method over a time span of 60 ns when considering
the topological transitions, and over 360 ns when
considering ratchet effects.
In previous work on chiral magnetic hopfions,
the dynamics were integrated
over a time span of 15 ns; however, here we
are focusing on topological transitions and ratchet effects,
and therefore require longer simulation times.

Our sample has parameters $J=1$ meV, $D=0.2J$, $K_V=0.01J$, and $K_{T,B}=5J$.
These parameters roughly correspond to the magnetic parameters of the chiral magnet MnSi
\cite{iwasaki_current-induced_2013, iwasaki_universal_2013, Wang19a}.

To ensure the stability of the hopfion, we first used the ansatz presented
in Knapman {\it et al.} \cite{Knapman24}
and performed a 60 ns integration with no SOT.
This integration gives a final ${\bf m}$ configuration
that we then employed as the initial condition for our
simulations.

\section{Results}

\begin{figure}
\centering
\includegraphics[width=0.9\columnwidth]{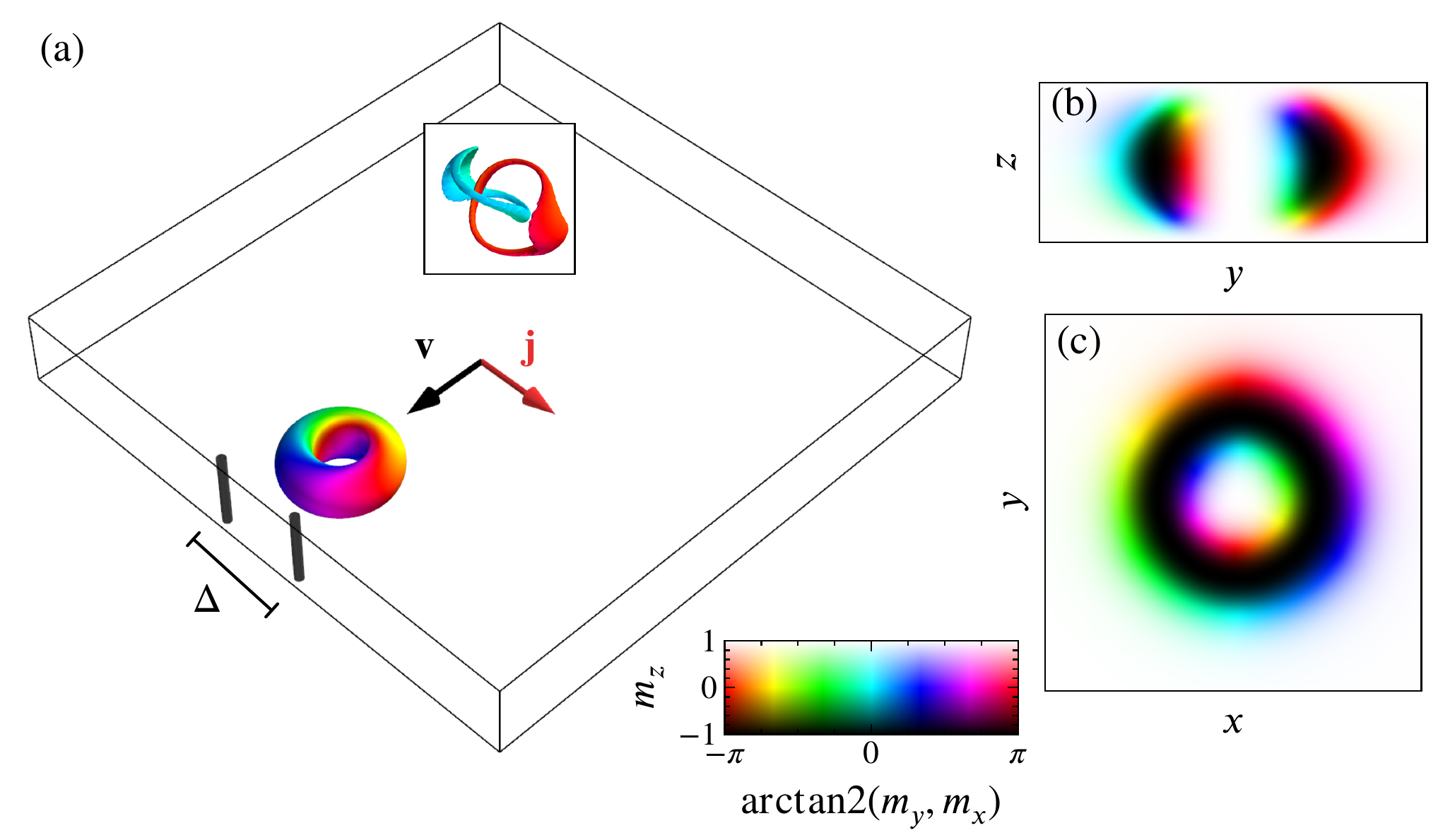}
\caption{
(a) 3D view of our system for a $Q_H = 1$ hopfion,
which is plotted by visualizing the $m_z=0$ isosurface.
The colors represent the $m$ direction given by the angle between
$m_x$ and $m_y$, as shown in the colorbar.
We place two column-like defects (thick black lines) with an increased PMA of $K_D$,
where $\Delta$ is the distance between the two columns.
The hopfion is subjected to dc driving
along ${\hat {\bf y}}$ produced by a current $j$.
This moves the
hopfion along ${\hat {\bf v}}={\hat {\bf x}}$ at a velocity $v$ towards
the center point between the two columns.
Inset: hopfion isosurfaces at $m_x=1$ (red) and $m_x=-1$ (cyan).
(b) $zy$ and (c) $xy$ cross-sections along the hopfion center.
The hopfion has a diameter of approximately $\xi=25$ nm.
}
\label{fig:1}
\end{figure}

We consider a $Q_{H} = 1$ 3D hopfion driven through a comb-like array of line defects that have a higher value of $K_{D}$ than the surrounding medium.
In Fig.~\ref{fig:1}(a), we show a 3D rendering of the
$m_z=0$ isosurface for the hopfion. The inset 
shows the $m_x=1$ and $m_x=-1$ isosurfaces,
highlighting the knot-like structure that is characteristic of
a hopfion with $Q_{H} = 1$.
A side view and a top view of the hopfion appear in
Fig~\ref{fig:1}(b) and Fig.~\ref{fig:1}(c), respectively.
In this case, the hopfion diameter is approximately 25 nm.
Arrows in 
Fig.~\ref{fig:1}(a) indicate the direction of the applied current $j$
as well as the direction ${\bf v}$ along which the hopfion moves.
The two black pillars are the columnar line defects, which are placed a distance $\Delta$ apart.
For different values of $j$, $\Delta$, and $K_D$, we apply the current and
wait for the system to evolve to a pinned
or sliding state or to transform
into a toron when the hopfion moves between the defects.
Due to the confinement inside the sample, 
the hopfion motion in the absence of the line defects
is similar to what was observed previously in Ref.~\cite{Wang19a},
where the hopfion moves without a Hall angle.

\begin{figure}
 \centering
 \includegraphics[width=0.9\columnwidth]{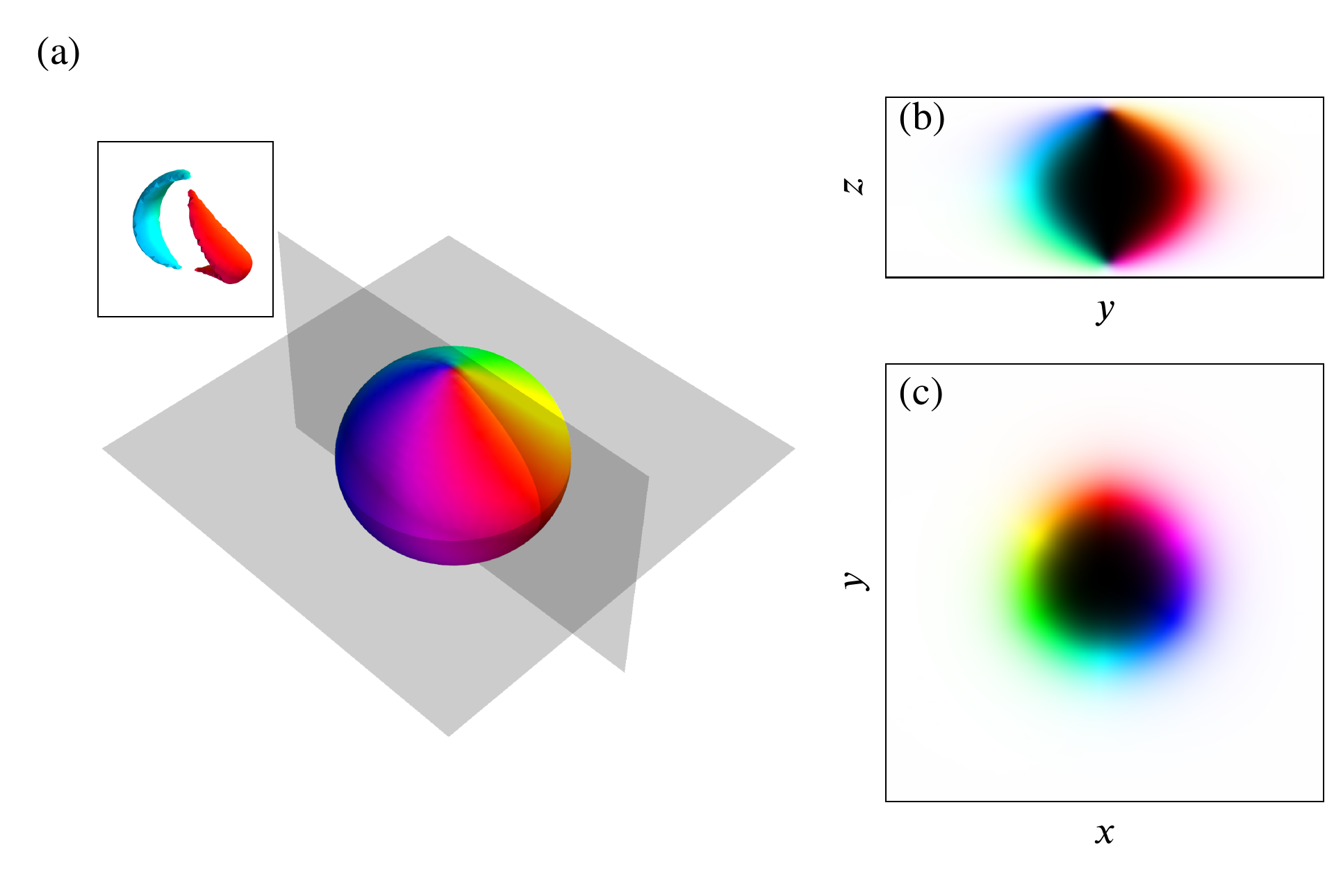}
 \caption{A dipole string texture or a toron with $Q_H=0$.
The colors represent the $m$ direction given by the angle between
$m_x$ and $m_y$, as in Fig.~\ref{fig:1}.
The hopfion topologically transitions to the toron texture
for certain combinations of $\Delta$, $K_{D}$ and $j$.
Inset: the $m_x=1$ (red) and $m_x=-1$ (cyan) isosurfaces.
(b) $zy$ (c) $xy$ cross-sections along the dipole string center.
The dipole string diameter is approximately $\xi=12$ nm.}
\label{fig:2}
\end{figure}

As a function of changing drive and defect parameters, we find three phases:
(i) a pinned phase where the current is so low that the hopfion
cannot move between the line defects and is pinned between them,
(ii) a sliding phase where the hopfion can distort and
pass between the defects, and 
(iii) a regime where the hopfion becomes so
strongly compressed upon moving between the defects that it transforms
into a dipolar string state
or a toron \cite{Muller20,Li20}
with  $Q_{H} = 0$, as shown in Fig.~\ref{fig:2}(a).
Here, the ring-like structure associated with
the $Q_{H} = 1$ hopfion in Fig.~\ref{fig:1} is missing,
and the plot of the $m_x=1$ and $m_x=-1$ isosurfaces in the inset
of Fig.~\ref{fig:2} indicate that 
there is no knotting.
The $yz$ or side view cross-section and the $xy$ or top view
cross-section of the toron appear in Figs.~\ref{fig:2}(b) and (c),
respectively.
The toron is close to $\xi=12$nm in diameter,
approximately half the size of
the hopfion.

\begin{figure}
 \centering
 \includegraphics[width=0.9\columnwidth]{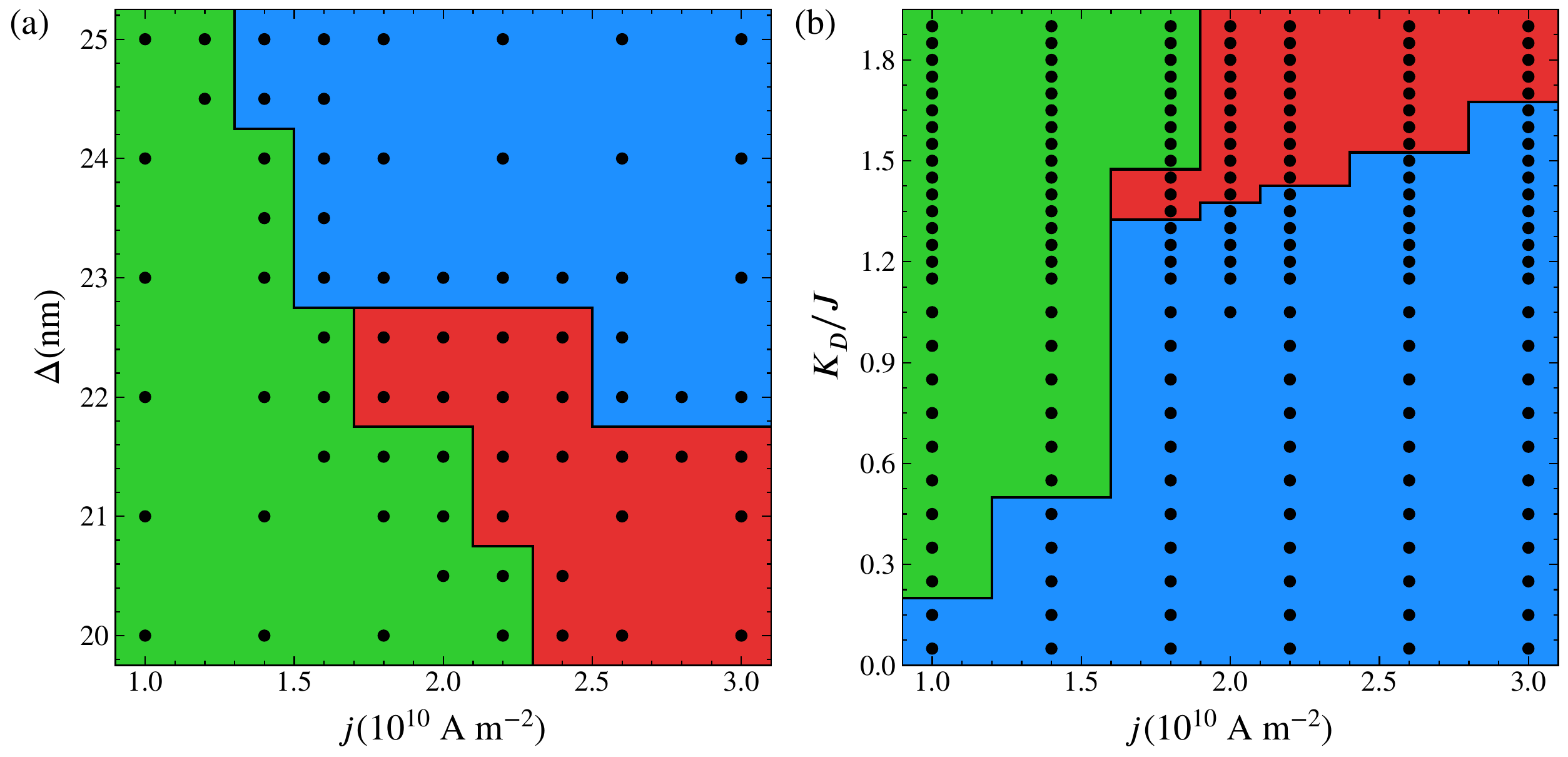}
 \caption{Phase diagrams for a hopfion interacting with
  a line defect structure as in Fig.~\ref{fig:1}(a),
  showing a pinned state (green),
  hopfion motion between the defects (blue), and a topological
  transition from a $Q_H=1$ hopfion to an
  unlinked $Q_H=0$ toron $Q_H=0$ (red).
  (a) Phases as a function of the
  spacing between the two columns $\Delta$ vs current $j$
  for a defect strength of $K_D=5J$.
  (b) Phases as a function of the defect strength
  $K_D/J$ vs current $j$ for a defect spacing
  of $\Delta=21$ nm.
 The black dots indicate the points at which simulation data was obtained.}
 \label{fig:3}
\end{figure}

In Fig.~\ref{fig:3}(a), we plot a phase diagram as a function of the
defect spacing $\Delta$ versus current $j$ in a system with a
defect strength of $K_D=5J$.
We find
a pinned phase, sliding motion of the hopfion
between the defects,
and a phase in which the $Q_H=1$ hopfion undergoes
a topological transition to a $Q_H=0$ toron.
The width of the pinned phase
grows with decreasing $\Delta$. 
When $\Delta > 23$ nm, the hopfion can depin into a moving hopfion
as the current increases. Since the hopfion diameter is
about 24 nm, the distortions induced in the hopfion by the defects as it
passes between them are reduced when
$\Delta > 23$ nm.
For $\Delta < 23$ nm, we find a pinned phase and a toron state;
however, when $j$ is sufficiently large,
the hopfion moves between the defects so rapidly 
that it does not deform strongly enough to transform into a toron,
and the moving hopfion phase reappears.
In previous work on systems with depinning,
it has been shown that the distortions
produced by the pinning on the moving textures is effectively
reduced at higher drives \cite{Reichhardt17a},
and this effect has been studied for skyrmions
in both particle based \cite{Reichhardt15}
and continuum models \cite{Koshibae19}.

The ability of the defects to transform the hopfion to a toron
is affected by the
defect strength.
In Fig.~\ref{fig:3}(b) we present a phase diagram as a function of
$K_{D}/J$ versus $j$ for fixed $\Delta = 21$ nm,
which is smaller than the hopfion width.
For $K_{D} < 1.3J$ the hopfion can depin into a moving hopfion state, while
for $K_{D} > 1.3J$ the hopfion transforms into a toron upon depinning, and
a larger $j$ must be applied in order to stabilize a moving hopfion.

\begin{figure}
\centering
\includegraphics[width=0.9\columnwidth]{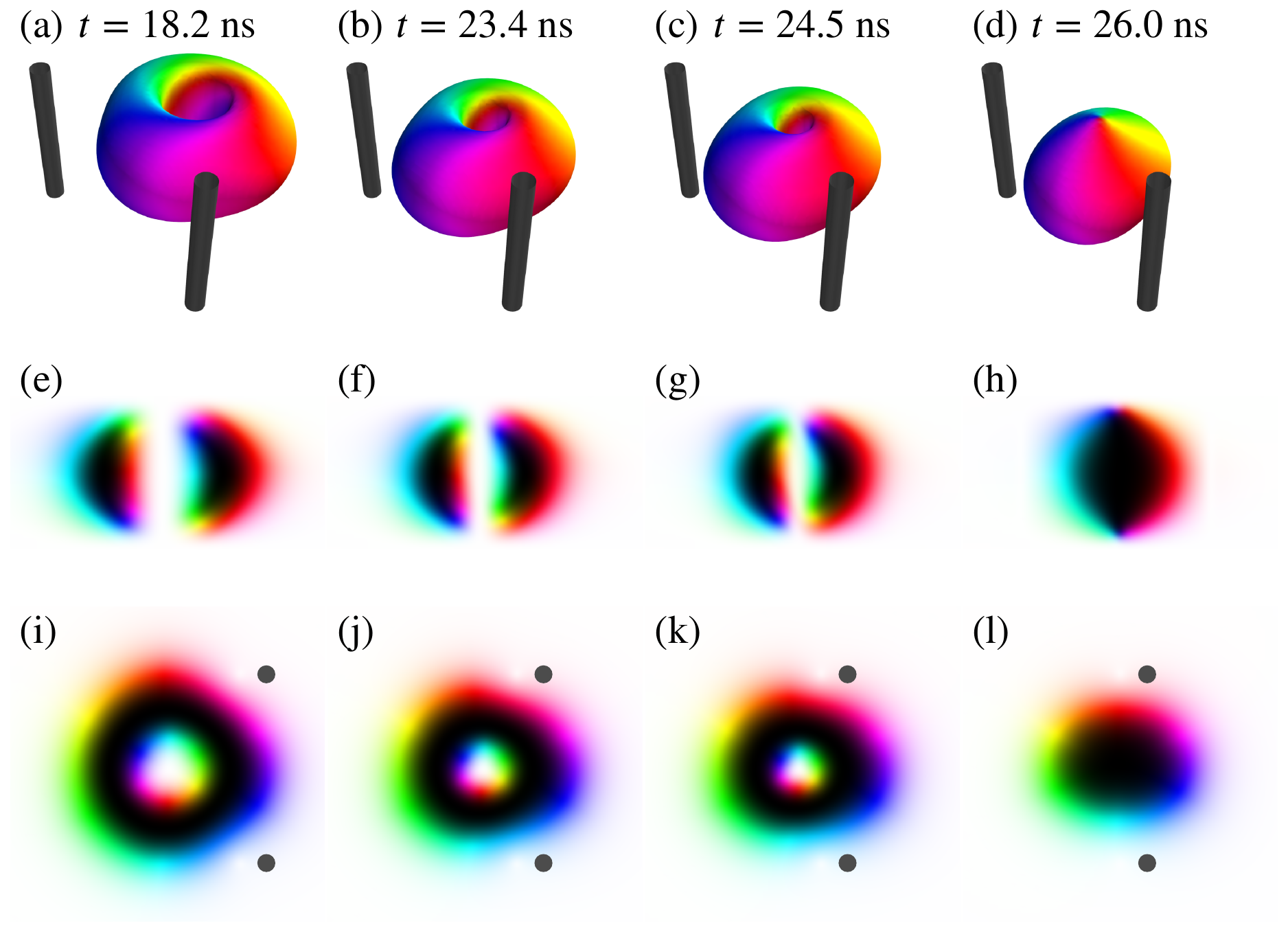}
\caption{Images of a hopfion topologically
transitioning from $Q_H=1$ to $Q_H=0$
upon interacting with the columnar defects.
The colors represent the $m$ direction given by the angle between $m_x$ and
$m_y$, as in Fig.~\ref{fig:1}.
Here $\Delta=21$ nm and $j=2.6\times10^{10}$ A m$^{-2}$.
(a,b,c,d) $m_z=0$ isosurfaces.
(e,f,g,h) The $yz$ cross-section.
(i,j,k,l) The $xy$ cross-section.
 }
 \label{fig:4}
\end{figure}

In Fig.~\ref{fig:4}, we illustrate the pinning-induced transition of the
hopfion into a toron
for the system from Fig.~\ref{fig:3}(a)
at $\Delta = 21$ nm, $K_{D} = 5J$, and $j=2.6\times10^{10}$ A m$^{-2}$.
Figure~\ref{fig:4}(a,b,c,d) 
shows 3D images of the $m_z = 0$ isosurface of the texture
along with the defect positions,
while Fig.~\ref{fig:4}(e,f,g,h) show the side view
and Fig.~\ref{fig:4}(i,j,k,l) show the top view
of the texture.
As the hopfion moves between the line defects, it becomes increasingly
compressed, and at $t = 26$ ns, it collapses into the $Q_{H} = 0$ state.
The images also show
that the toron is smaller than the hopfion.

\begin{figure}
\centering
\includegraphics[width=0.9\textwidth]{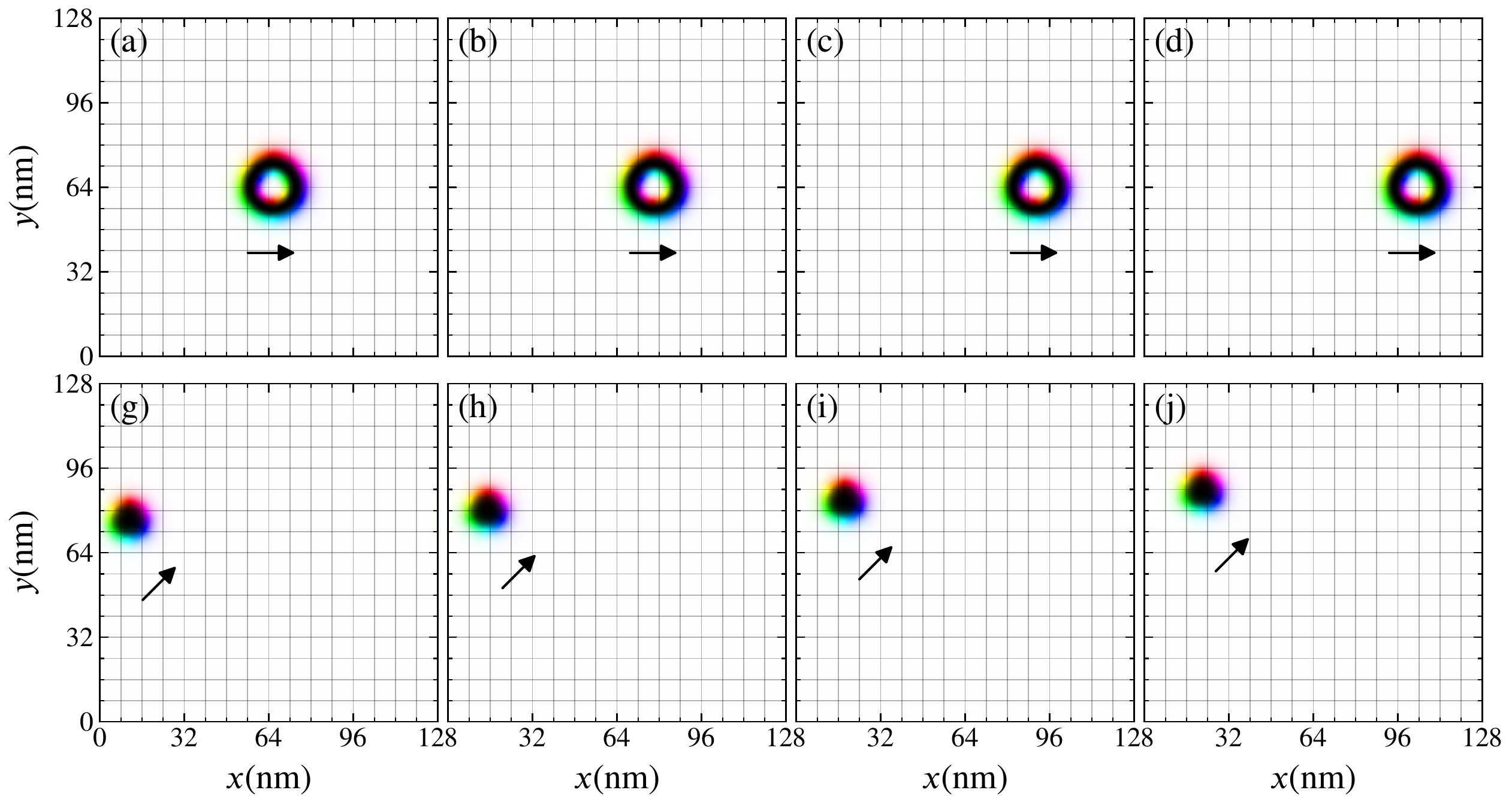}
\caption{Images of freely moving textures away from the influence
of the defect lines in the system from Fig.~\ref{fig:4}
with $\Delta=21$ nm and
$j=2.6\times10^{10}$ A m$^{-2}$ where a hopfion to toron transformation
occurs.
The colors represent the $m$ direction given by the angle between $m_x$ and
$m_y$, as in Fig.~\ref{fig:1}.
(a-d) Hopfion motion, with no Hall angle, $\theta=0^\circ$,
before the transformation at
(a) $t=0$ ns,
(b) $t=5.1$ ns,
(c) $t=9.6$ ns, and
(d) $t=14.1$ ns.
(g-j) Dipole string, or toron, motion, with a Hall angle of
$\theta=43^\circ$
after the transformation at
(a) $t=40.8$ ns,
(b) $t=45.3$ ns,
(c) $t=49.8$ ns, and
(d) $t=54.3$ ns.
The velocity of the hopfion 
is roughly $v=2.7$ m s$^{-1}$, and that of the
dipole string is roughly $v=1.1$ m s$^{-1}$.
Arrows indicate the direction of motion.}
\label{fig:5}
\end{figure}

In Fig.~\ref{fig:5}(a-d), we show the top view of a moving hopfion at
$j=2.6\times10^{10}$ A m$^{-2}$ in the sample from Fig.~\ref{fig:4} where
a hopfion to toron transition occurs. In the images, the hopfion is
not close to the columnar defects and is not affected by them.
The hopfion moves at a velocity $v=2.7$ m s$^{-1}$, without a Hall angle.
Figure~\ref{fig:5}(g-j) shows the top view of the moving toron
away from the columnar defects after the transition.
The toron travels at a velocity of $v=1.1$ m s$^{-1}$
with a Hall angle of approximately $\theta=43^\circ$.

\begin{figure}
\centering
\includegraphics[width=\textwidth]{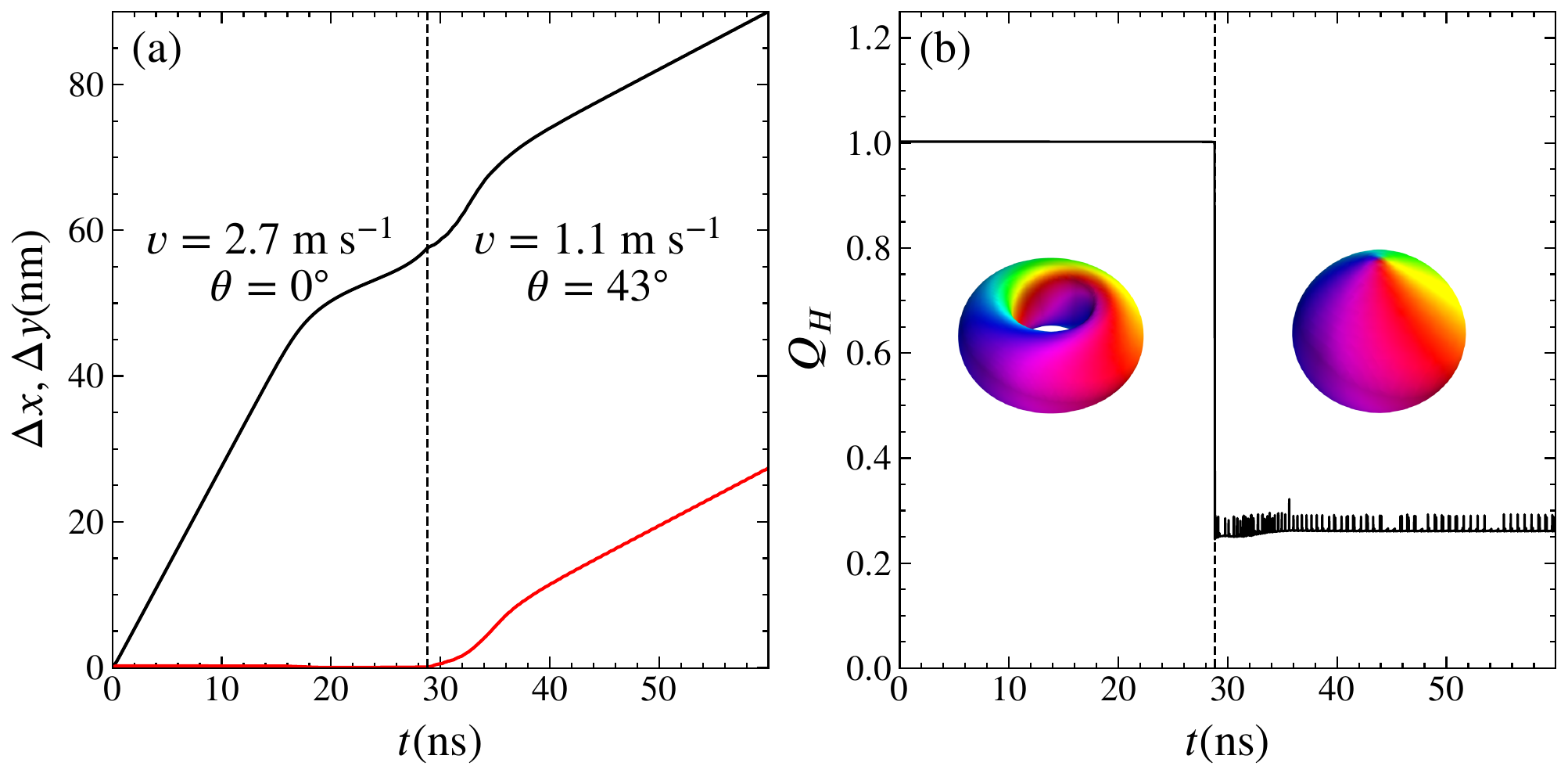}
\caption{(a) Displacements $\Delta x$ (black)
and $\Delta y$ (red) versus time for the system from Fig.~\ref{fig:4}
with $j=2.6\times10^{10}$ A m$^{-2}$, $\Delta=21$ nm, and $K_D=5J$,
where a transition from a hopfion to a toron occurs as the hopfion
passes between the defects.
The hopfion travels at a higher velocity
of $v=2.7$ m s$^{-1}$ and with no Hall
angle, whereas the dipole string moves
at a slower velocity of $v=1.1$ m s$^{-1}$ with
a Hall angle of $\theta=43^\circ$.
(b) The corresponding Hopf index $Q_H$ versus time.
The oscillations in $Q_H$ that appear after the transition are numerical
artifacts.
In each panel, a dashed line marks the moment at which the
hopfion to toron
transition occurs.
}
 \label{fig:6}
\end{figure}

In Fig.~\ref{fig:6}, we show the displacements $\Delta x$ (black)
and $\Delta y$ (red) versus time for the system in Fig.~\ref{fig:4}
with $j=2.6\times10^{10}$ A m$^{-2}$, $\Delta=21$ nm, and $K_{D} = 5J$,
where the hopfion transforms to a toron near $t = 28ns$ upon
interacting with the line defects.
The hopfion moves with a higher velocity of $v=2.7$ m s$^{-1}$,
visible
as the larger slope in $\Delta x$,
while $\Delta y = 0.0$ for the hopfion, indicating
that there is no Hall angle.
When the hopfion begins to interact with the defects
near $t = 20ns$, there is a dip in $\Delta x$ as the texture slows down.
After the transition, the toron moves at $v=1.1$ m s$^{-1}$ with
a Hall angle of $\theta=43^\circ$,
which appears as the onset of finite $\Delta y$ and a reduced
slope of $\Delta x$.
In Fig.~\ref{fig:6}(b) we plot the
corresponding Hopf index $Q_H$ versus time.
For $t<28$ ns, $Q_H = 1$ and a hopfion is present,
while at later times, $Q_H$ drops in the toron state.
The oscillations in $Q_H$ that appear after the transition are numerical
artifacts.

Since the line defects are able to pin the hopfion,
this suggests that certain types
of defects could also guide the hopfion motion.
In addition, if an asymmetric arrangement of defects were
introduced,
the system could exhibit a ratchet effect.
In a rocking ratchet,
an ac drive applied to a particle on an asymmetric substrate produces
a net dc drift motion.
Rocking ratchets have been realized for
overdamped superconducting vortices interacting
with asymmetric substrates under ac driving
\cite{Lee99,Villegas05,deSouzaSilva06a,Yu07,Luo07,PerezdeLara11}.
Ratchet effects have also been observed
for skyrmions interacting with asymmetric substrates;
however, in the skyrmion system,
the Magnus force provides
additional ways for the texture to be rectified
\cite{Reichhardt15a,Gobel21,Souza21}. 
Since hopfions do not have a Hall angle, their ratcheting motion would
be expected to be comparable
to that of the overdamped superconducting vortex systems.
A hopfion ratchet would then
represent a realization of a ratchet for a 3D object.

\begin{figure}
\centering
\includegraphics[width=\textwidth]{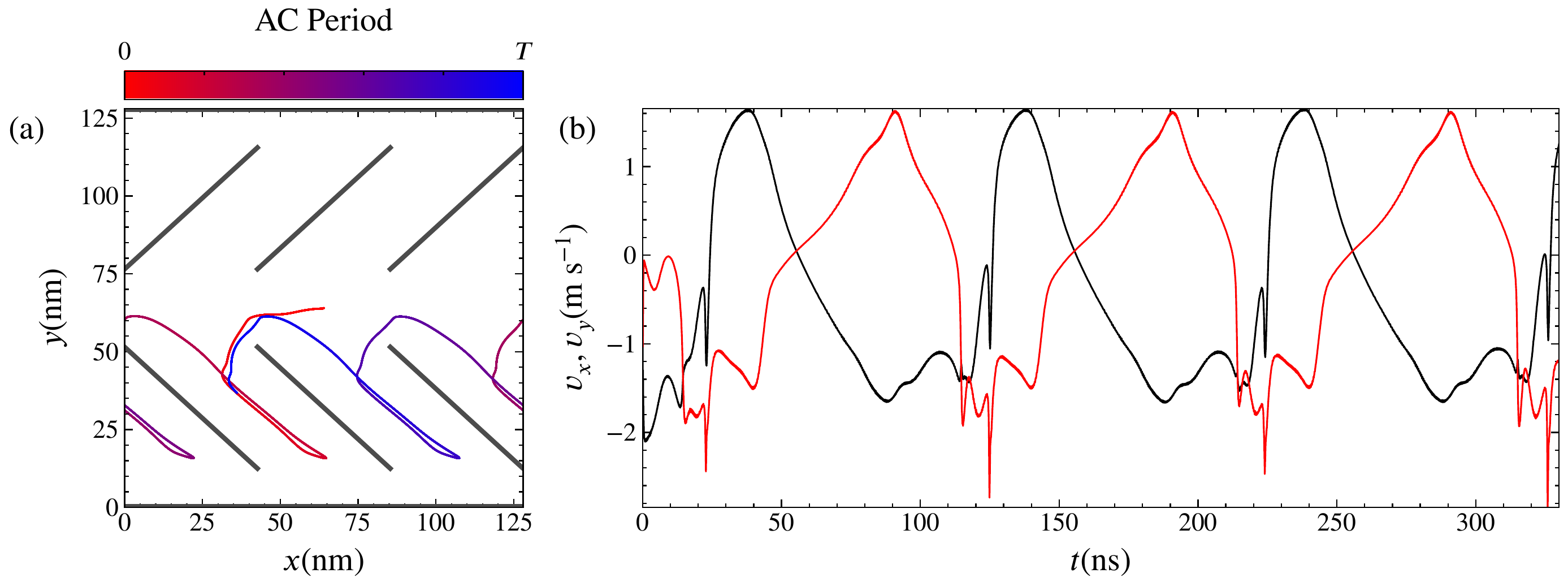}
\caption{(a) Image of the trajectory of a hopfion, colored according to
the progression of the circular ac driving period, moving through
a herringbone planar defect array pattern (grey lines)
in a sample with $K_D=0.5J$.
(b) The hopfion velocity $v_x$ (black) and $v_y$ (red) vs time for
the system in (a).
An animation showing the hopfion ratchet motion is available in the Supplemental Material \cite{supplemental}.
 }
 \label{fig:7}
\end{figure}

In Fig.~\ref{fig:7}(a) we show a top view of a substrate
containing a herringbone planar defect array
with anisotropy of $K_D=0.5J$.
We place a hopfion in the center region near $x = 75nm$ and
apply a circular ac drive given by
${\bf j}_\text{AC} = j_\text{AC}\sin(2\pi f t){\hat{\bf x}} - j_\text{AC}\cos(2\pi f t){\hat{\bf y}}$,
with $j_\text{AC} = 2\times10^{10}$A m$^{-2}$
and frequency $f = 0.01$GHz.
The lines show the hopfion trajectory, where we change the
color for each successive ac cycle to show that the
hopfion is translating along the negative $x$-direction corresponding
to the easy flow direction of the substrate asymmetry.
Figure~\ref{fig:7}(b) shows the hopfion velocity $v_x$ and $v_y$
versus time for
the system in Fig.~\ref{fig:7}(a).
The texture translates by one substrate lattice site 
per cycle in the negative $x$ direction.
In this case, we must apply a circular ac drive in order to break enough
symmetry to obtain a ratchet effect;
however, the finite Magnus force of
the toron could produce an additional circular motion so that 
an ac drive applied along only one axis could be sufficient to produce
ratchet transport.
Other defect geometries
could permit the ratcheting motion of the hopfions
under uniaxial ac driving.

\section{Discussion}

Our results indicate that hopfions can interact with and be controlled by structured defect arrays. We considered defects created by varying the perpendicular magnetic anisotropy; however, defects could be also generated by varying other parameters of the sample. The line defects we study could be fabricated by a variety of irradiation techniques, such as a focused ion beam pattern. Alternatively, the surface could be patterned with magnetic dots or via surface etching.
Although we studied only a single hopfion,
we expect that additional dynamical behaviors could arise
for multiply interacting hopfions or for hopfions with higher Hopf indices,
where there could be transitions from higher to lower indices.
We found that the hopfion transforms to the dipolar string texture upon
interacting with defects, but
it would also be interesting to see if the reverse
transformation could occur,
or if a nanostructured pattern could be used to nucleate hopfions in
a controlled fashion.
In our system, the hopfions are confined, and most experiments with
line defects would likely fall into the same type of
confined geometry, but it would also be interesting to study
hopfion dynamics in bulk systems
where it is known that there can be additional 3D dynamics \cite{Liu22}.
We also considered only line-like and planar defects,
but surface defects or defects in the bulk would also be
interesting to explore.
This topological transition is akin to that observed for a skyrmionium
turning into a skyrmion upon passing
through a constriction.

\section{Conclusion}

Using atomistic simulations, we have examined the dynamics of driven hopfions
interacting with line-like defects.
As we vary the driving,
the distance between the defects,
and the defect disorder strengths,
we find three phases: a pinned state,
a sliding hopfion phase where there
is a critical depinning drive and the hopfions can
move between the defects by distorting, and a state
where the hopfion transforms into a dipolar string or toron that
is half the size of the hopfion
and moves at a lower velocity for the same applied current.
For disorder strengths and defect spacings
at which the hopfion transforms into a toron
at depinning, we find that if the drive is increased sufficiently far
above the depinning transition, a moving hopfion
can be stabilized. The hopfion moves without a Hall angle, but the toron
moves with a finite Hall angle.
When we introduce
an asymmetric arrays of planar defects,
a hopfion ratchet effect can be realized under circular ac driving.
Our results suggest that the position and dynamics of
hopfions can be controlled with judiciously constructed defect arrays.
This opens the possibility of a variety of applications utilizing hopfions,
such as memory elements or new types of computing devices.
In addition to magnetic textures, our results are 
also relevant to hopfions in ferroelectrics or
in liquid systems under fluid flow conditions. 

\section{Acknowledgments}
This work was supported by the US Department of Energy through the Los Alamos National Laboratory. Los
Alamos National Laboratory is operated by Triad National Security, LLC, for the National Nuclear Security
Administration of the U. S. Department of Energy (Contract No. 892333218NCA000001). 
J.C.B.S  and N.P.V. acknowledge funding from Fundação de Amparo à Pesquisa do Estado de São Paulo - FAPESP (Grants J.C.B.S 2023/17545-1 and 2022/14053-8, N.P.V 2024/13248-5).
We would like to thank FAPESP for providing the computational resources used in this work (Grant: 2024/02941-1). 

\bibliography{mybib}

\end{document}